# An Identification of Learners' Confusion through Language and Discourse Analysis

Thushari Atapattu, Katrina Falkner, Menasha Thilakaratne, Lavendini Sivaneasharajah, Rangana Jayashanka

*Abstract*— The substantial growth of online learning, in particular, Massively Open Online Courses (MOOCs), supports research into the development of better models for effective learning. Learner 'confusion' is among one of the identified aspects which impacts the overall learning process, and ultimately, course attrition. Confusion for a learner is an individual state of bewilderment and uncertainty of how to move forward. The majority of recent works neglect the 'individual' factor and measure the influence of community-related aspects (e.g. *votes*, *views*) for confusion classification. While this is a useful measure, as the popularity of one's post can indicate that many other students have similar confusion regarding course topics, these models neglect the personalised context, such as individual's affect or emotions. Certain physiological aspects (e.g. *facial expressions, heart rate*) have been utilised to classify confusion in small to medium classrooms. However, these techniques are challenging to adopt to MOOCs. To bridge this gap, we propose an approach solely based on language and discourse aspects of learners, which outperforms the previous models. We contribute through the development of a novel linguistic feature set that is predictive for confusion classification. We train the confusion classifier using one domain, successfully applying it across other domains.

## I. INTRODUCTION

CONFUSION is an emotional and cognitive state [1] often experienced during learning, in particular when learning new or complex information [2]. Learners are more likely to exhibit confusion when new knowledge integration conflicts with existing knowledge structures. As found in the authors' previous work [3], lack of support for knowledge organisation in educational materials impedes the integration of new knowledge with prior knowledge, which contributes towards confusion. Although inferred as 'negative' for learning progression, a certain degree of confusion is considered constructive and can promote cognitive engagement [4, 5].

The *model of affect dynamics* describes the transition of emotions during learning tasks using four states – *flow/engagement* ↔ *confusion* ↔ *frustration* ↔ *boredom* [4, 6, 7]. When a state of confusion is triggered, *timely interventions* [8-11] are crucial to prevent the transition of the learner's emotional state to frustration, and eventually, to boredom. Building on this, D'Mello *et al.* [4] conceptualised a minimum level of constructive confusion and maximum level of adverse confusion as the *zone of optimal confusion* [12, 13]. At the lower boundary of the zone of optimal confusion, the learner can either resolve the confusion and return to *flow/engagement* [12, 13] or transfer to the bewildered state with the uncertainty of future learning pathways [14]. Prior research related to the impact of emotions (e.g. confusion) also found the potential for course attrition in online and distant learning, in particular, Massively Open Online Courses (MOOCs) [14-17]. Further, learners that persist despite confusion are more likely to proceed with misconceptions or incorrect knowledge structures.

According to Lehman *et al.* [5], the zone of optimal confusion is related to 'individual' aspects such as age, motivation, personality, and confidence [2]. Therefore, the design of instructor interventions to aid confusion will vary from one learner to another. Through an empirical study, Yang *et al.* [14] also suggest that confusion is a behaviour associated with individuals. In their recent work, Lodge *et al.* [18] proposed a framework for understanding confusion in learning. According to Lodge *et al.* [18], confusion can be either productive or unproductive depending on a range of variables including the design of learning tasks and activities, individual differences (e.g. prior knowledge, self-efficacy, self-regulation), and design and timeliness feedback.

Arguel *et al.* [2] summarised techniques to detect learners' confusion in a personalised context, in particular, Intelligent Tutoring Systems (ITSs), including facial expressions [19], conversational clues [20], and skin conductance [21]. While these techniques are useful in the context of small to medium classrooms, they are challenging to adopt in the context of online and distant learning, as they rely on equipment such as tracking devices which constrains deployment at scale. As a consequence, new techniques have been explored to achieve scale, including the use of clickstream data and forum participation analysis [2, 8, 14, 22-24].

Although the theory underpinning confusion emphasises individual aspects [6, 12, 13, 18], the majority of recent work [8, 9, 23] neglects the 'individual' factor and measure the influence of community-related aspects (e.g. how a community reacts to an individual's post using *votes*, *views*) to classify learners' confusion. While this is a useful measure, as the popularity of one's post can indicate that many other students have similar confusion regarding course topics, assessments, etc., these models demonstrate drawbacks within the personalised context. Firstly, participants across the globe partaking from different geographical locations, time zones, with varied expectations on studying pace and knowledge/expertise level, it is likely that the confused learner will reach the zone of optimal confusion by the time they are able to receive personalised assistance or support from the course community and/or instructors. Therefore, real-time confusion identification is challenging with the use of community-related metadata. Secondly, these models are likely to misclassify 'confused' posts if they don't receive satisfactory reactions from the community, causing some students to be left behind despite instructor interventions.

To bridge this gap and detect confusion independently from community-related factors within the MOOC context, we investigate an approach solely based on the language and discourse aspects of discussion posts. D'Mello & Graesser [7] explored a similar study in classifying 'emotions' including confusion between student-tutor pairs in the context of ITSs. They emphasise the importance of 'conversational dynamics' of both student and tutor to predict emotions. While sharing the similar intentions as related works [8, 14, 23, 25], we hypothesise that the "*learners' confusion can be effectively understood solely based on the way they express the language irrespective of the subject domain*".

To test our hypothesis, we utilise the Stanford MOOC posts dataset [8, 26], experimenting using forum posts from three different disciplines (i.e. Education, Medicine, and Humanities). The dataset consists of textual posts and their corresponding degree of confusion ranging from 1 (extremely knowledgeable) to 7 (extremely confused). We categorise the posts into two classes as *confused* (score $>= 4$) and *non-confused* (score $< 4$).

Accordingly, our first research question RQ1 asks, *What language and discourse features demonstrate a significant difference between confused and non-confused forum posts?*. To answer this question, we conduct a Multivariate Analysis of Variance (MANOVA), using linguistic features as dependent variables and the post category (e.g. confused) as an independent variable. Knowing the answer to this question, our second research question RQ2 investigates, *Can we build machine learning models for confusion classification solely based on linguistic features?*. We use these models to answer our third research question RQ3, *Can we successfully validate the linguistic-based machine learning models for confusion classification on unseen domains?*.

Primarily, our work contributes to producing a novel set of predictive linguistic features for confusion classification irrespective of the subject domain. Further, we built machine learning models that outperform the previously developed algorithms with the use of an identical dataset [8, 23, 25]. As our approach is independent of community-related features and solely based on the way learner expresses the language, the 'early' identification of confusion is feasible without the influence from the community (e.g. views, votes).

## II. RELATED WORK

Previous related research can be categorised as research on confusion detection using external devices, sensor-free confusion classification, and feature space development for confusion classification.

Arguel *et al.* [2] summarised three methodologies, i.e. self-report, behavioural, and physiological techniques to detect learners' confusion in a personalised context, in particular, ITSs, including facial expressions [19], postures [6], conversational clues [20], and skin conductance [21]. A comprehensive analysis of the literature on confusion detection using behavioural and physiological techniques is included here [2].

Despite the promise of facilitating the collection of individual's traces (e.g. self-report, physiological) [2], the research on confusion detection using external devices constrains deployment at scale. Therefore, the efforts of sensor-free techniques applicable to online learning are discussed here.

Some original works in identifying confusions in MOOC discussions include the study by Agrawal *et al.* [8] who also contributed to the development of Stanford MOOC posts dataset, used in our study and other works [9, 23, 25] (more details about the dataset is included in Section 3.A). Agrawal *et al.* [8] implemented a system called YouEDU that identifies the confusion in the posts using Bag-of-Words[1] and post metadata (e.g. *votes, post position*) features and recommended supplementary video snippets accordingly.

Preliminary work by Bakharia [25] utilises the Stanford dataset to predict confusion, urgency, and sentiments of MOOC forum posts. Using a Support Vector Machine (SVM) with an RBF kernel, Bakharia achieved over 70% accuracy for confusion classification in individual courses. However, due to the incorporation of domain-specific language into her model, her model lacks generalisability across domains.

Zeng *et al.* [23] built a model for confusion classification using the Stanford dataset [26] which claimed to outperform the previous algorithms. They utilised a content-related (e.g. *post length, readability score*) and community-related (e.g. *votes, reads*) features and achieved over 80% of classification accuracy for individual domains (e.g. education). Additionally, they demonstrated a classification accuracy over 65% across domains, i.e. training on one domain and testing on another (more information about the comparisons with our models is included in Fig. 1 & 2).

Although not directly mentioning 'confusion', quite a few studies contribute to the related literature. Omaima *et al.* [9] utilised the Stanford dataset to build machine learning models for *urgency* classification. Their work contributes to expedite instructor interventions by emphasising the level of criticality of the posts. Corrin *et al.* [27] and Cross *et al.* [28] classify the help-seeking behaviour in online communities (e.g. StackExchange). Prediction of instructor intervention using content- and thread-related features is also reported in Chaturvedi *et al.* [11].

---

[1] https://en.wikipedia.org/wiki/Bag-of-words_model

As outlined in the Introduction, the theory underpinning confusion emphasises the importance of 'individual' aspects [6, 12, 13], compared to the influence of community-related structural aspects (e.g. *votes, reads*) being used for confusion classification in prior studies [8-11, 23]. A feature ablation study by Zeng *et al.* [23] also found that community-related features have less impact on the classifier performance whereas *unigram* features and *question mark* demonstrated the highest impact.

To address this gap, limited research studies focus on traces left by individuals without relying on external devices, including self-report, clickstream patterns/log files [14, 22, 24, 29], and more importantly, language and discourse features [7, 14, 20]. Baker *et al.* [22] built models for affect classification using students' interactions (i.e. log files) within a Cognitive tutor for Algebra.

A study by Yang *et al.* [14, 29] integrated a combination of course activity patterns (e.g. video watching) and discussion contents into their confusion classification model. Their dataset was constructed using two Coursera MOOCs for Algebra and Micro Economics. Although they have achieved a classification accuracy over 70% using a combination of features such as *click patterns, language*, their linguistic-only features demonstrate a relatively low classification accuracy (~60%). Due to the diversity of course designs including self-paced, offline access to materials without data traces, community-centric models [30], and allowing anonymity, relying on clickstream data for confusion classification presents some constraints.

Thus, our work investigates the potentials of using language and discourse features of discussion posts for confusion classification. D'Mello & Graesser [7] presented a related study in classifying 'emotions' including confusion using a language and discourse analysis between student-tutor pairs within ITSs. However, authors emphasise the importance of 'conversational dynamics' of both student and tutor to predict emotions which is less applicable to MOOCs context.

*A. Feature space development for confusion classification*

Over the past decade, the majority of related studies focused on the development of a rich feature set which has the potential to predict confusion. We summarise the features utilised so far for confusion/urgency classification below.

TABLE I
FEATURE SPACE FOR CONFUSION/URGENCY CLASSIFICATION

| Feature category/Study | [8] | [25] | [14] | [23] | [7] | [22] | [9] |
|---|---|---|---|---|---|---|---|
| Unigram | ✓ | ✓ | ✓ | ✓ | | | ✓ |
| Content | | | | ✓ | | | |
| Language | | | ✓ | | ✓ | | ✓ |
| Question | | | ✓ | ✓ | | | |
| Activity | | | ✓ | | | ✓ | |
| Community | ✓ | | | ✓ | | | ✓ |
| Post metadata | ✓ | | | | | | ✓ |

**Studies**: Agrawal *et al.* [8], Bakharia [25], Yang *et al.* [14], Zeng *et al.* [23], D'Mello & Graesser [7], Baker *et al.* [22], Omaima *et al.* [9];

**Feature examples in categories**:
- Unigram – Bag-of-words
- Content-related – post length, readability score, topicality
- Language – features (e.g. pronouns, sentiments) extracted from the tools like LIWC [31], Coh-metrix [32]
- Question – question mark, question stem (e.g. why, how)
- Activity – clickstream data (e.g. video watching)
- Community – votes, reads/views
- Post meta data – anonymous or not, poster's grade in the class, post position/type (e.g. new thread or reply comment)

III. RESEARCH STUDY DESIGN

Based on the theories and empirical studies [4, 6, 13] and the gaps identified in the prior studies [8, 23, 25], we argue that confusion is an affect related to 'individuals'. Therefore, we focus on identifying confusion solely based on the way learner expresses the language. Hence, we aim to answer the following research questions;

**RQ1**: What language and discourse features demonstrate a significant difference between confused and non-confused posts?

**RQ2**: Can we build machine learning models for confusion classification solely based on linguistic features?

**RQ3**: Can we successfully validate the linguistic-based machine learning models for confusion classification on unseen domains?

*A. Dataset*

Our dataset[2] for this study was obtained from the Stanford MOOC posts [8, 26]. It consists of eleven Stanford University public online courses from three domains (i.e. Humanities - HM, Education - EDU, and Medicine - MED). The courses contain a combination of topics such as Women's Health (HM), Statistics (MED), Scientific Writing (MED), and Mathematics (EDU). The dataset contains approximately 30,000 anonymised forum posts. This dataset is available for academic studies upon completing a required ethics component. Each forum post was coded by three independent coders on six dimensions: *question, answer, opinion, sentiment, confusion*, and *urgency*. The procedures on anonymisation, coding, and the creation of gold standard dataset are included on their website [26].

Our study considers only the textual posts and their corresponding degree of 'confusion' ranging from 1 (extremely knowledgeable) to 7 (extremely confused). Posts that received the score of '4' are considered as 'neutral'. Primarily, we consider neutral posts as confused to reduce false positives. We categorise the posts into two classes as

---
[2] https://datastage.stanford.edu/StanfordMoocPosts/

*confused* (score >= 4) and *non-confused* (score < 4). Related studies that experimented the same dataset either disregard neutral posts [23] or consider 'neutral' posts as confused [25]. To allow comparison with baselines, we present results with and without neutral posts (Fig. 1). Table 2 shows our corpus statistics.

TABLE 2
CORPUS STATISTICS

| Course | | Education | Humanities | Medicine | Total |
|---|---|---|---|---|---|
| Confused | Confused | 638 | 2236 | 1588 | 4462 |
| | Neutral | 2515 | 6015 | 6736 | 15266 |
| Non-confused | | 6690 | 1337 | 1563 | 9590 |
| Total | | 9843 | 9588 | 9887 | 29318 |

*B. Methods*

To accomplish our primary aim of confusion classification, i.e. 1) solely based on linguistic features, and 2) outperform the existing algorithms, we adopt quantitative analysis, using a combination of;
- automated language analysis,
- statistical analysis, and
- the building of machine learning models as our methodology.

For language analysis, our focus is to extract language and discourse features based on; 1) to what extent we could reuse the linguistic indices already being implemented through NLP tools, and 2) what other useful linguistic clues that could convey the confused state. We further categorise the probable confused linguistic clues as 1) 'direct' confusion expressions, and 2) expressions that could 'indirectly' convey confused state.

*1) Automated language analysis*

We utilise Sentiment Analysis and Cognition Engine (SEANCE) [33] for the analysis. SEANCE includes over 250 indices related to sentiments and emotions such as;
- positive sentiments (e.g. *love, like*),
- negative sentiments (e.g. *dislike, hate*),
- neutral sentiments,
- arousal words (e.g. *impatient*),
- positive emotions (e.g. *pleasure, enjoyment*), and
- negative emotion (e.g. *anger, fear*).

LIWC [31] has been widely used in similar studies [7, 9, 14]. Therefore, we extend their implications to our design. Further, Coh-metrix [32], which is held to be the most sophisticated tool grounded with theories of text and discourse comprehension (e.g. *cohesion, narrativity*), has also been used in a previous analysis [7]. However, Coh-metrix lacks support for offline corpus analysis and batch processing.

We also utilise Simple Natural Language Processing tool (SiNLP) [34]. SiNLP returns shallow (e.g. *number of words*) as well as deep language features (e.g. *lexical diversity, connectives*).

SiNLP also allows user-defined language categories. We define 'direct' confusion expressions as follows;
- negations (e.g. *couldn't, do not*),
- question-related features (e.g. question mark, question stem such as *what, how*, question bi-grams such as *can someone, what if*),
- confusion expressions (e.g. *exhaust, don't understand*),
- expressions related to 'incompleteness' (e.g. *missing*)
- error (e.g. *wrong, incorrect*), and
- problem solving (e.g. *problem, issue*).

Conversely, learners who do not directly express their confusion might relate it to pedagogy, as gratitude, or as pronouns. We define 'indirect' confusion expressions [7, 9, 28, 29] as follows;
- pedagogy-related (e.g. *lecture, class*),
- gratitude/politeness (e.g. *appreciate, please*),
- pronouns (e.g. *I, you*),
- determiners (e.g. *this, these*),
- opinion (e.g. *I believe, probably*), and
- future words (e.g. *will, might*).

Table 4 provides further examples from feature categories. SÉANCE[3] and SiNLP[4] can be installed locally and support batch processing.

*2) Statistical analysis*

Our method for statistical analysis utilises a Multivariate Analysis of Variance (MANOVA) (results are reported in Section 4.A). MANOVA is used as a common statistical test when multiple independent variables are present (i.e. multiple linguistic features in our case). Prior to the analysis, we test and ensure that the assumptions of MANOVA (e.g. normality of each of the dependent variables using Shapiro-Wilk test, homogeneity of variance using Levene's test, and multicollinearity) are not violated. Our focus of this analysis is to select significant language and discourse features for confusion classification, and hence, design a feature space.

*3) Building of machine learning models*

We utilise machine learning techniques to build predictive models for confusion classification. For this, we utilise significant language and discourse features as predictive variables and binary classification algorithms (e.g. Naïve Bayes, Random Forest) to build models (see results in Section 4.B).

IV. RESULTS

Table 3 reports the descriptive statistics of the Education domain. The choice of Education domain for reporting descriptive statistics was arbitrary (Note: similar descriptive statistics are available for other domains).

---

[3] https://www.linguisticanalysistools.org/seance.html
[4] https://www.linguisticanalysistools.org/sinlp.html

TABLE 3
DESCRIPTIVE STATISTICS OF EDUCATION

| Group | Posts | Sentences/post Mean (SD) | Words/post Mean (SD) | Words/sentence Mean (SD) | Letters/word Mean (SD) |
|---|---|---|---|---|---|
| Confused | 3153 | 3.47 (3.06) | 49.73 (49.04) | 14.63 (8.43) | 4.69 (2.47) |
| Non-confused | 6690 | 4.40 (3.73) | 72.41 (62.01) | 17.40 (8.60) | 4.65 (1.26) |
| Total | 9843 | | | | |

## A. Statistical analysis

To address our first research question (RQ1), we conducted a Multivariate Analysis of Variance (MANOVA), using linguistic features as dependent variables and the post category (i.e. confusion/non-confused) as an independent variable to measure any significant differences in language between the post categories.

We extracted 301 linguistic indices from the NLP tools. These were checked against assumptions of MANOVA. Indices which lacks normal distribution and exhibits multicollinearity ($r > 0.9$) were eliminated. The results indicated a statistically significant difference in language between groups. From these variables, we eliminate whose significance might be due to chance using Benjamini-Hochberg procedure [35]. Table 4 lists the significant variables with the *largest effect sizes* that were retained in the analysis within the Education domain. We report MANOVA results using the Education domain since our classifier achieved the best performance within the Education domain (Note: full list of predictive features will make available in public upon the work accepted for publication).

MANOVA results were fairly consistent across domains. In other words, language and discourse variables that are significant between confused and non-confused posts (e.g. *number of words, positive sentiments*) are fairly consistent between the three domains. Therefore, the language variables listed in Table 4 are the significant variables within all three domains (Note: F and are $\eta^2$ slightly different).

Conversely, Table 5 reports the remaining significant language variables that were not demonstrated significance within the Education domain. Similar to Table 4, we report results (i.e. F and $\eta^2$) of Humanities domain, however, these significant language variables are consistent within the Medicine domain (Note: F and are $\eta^2$ slightly different).

TABLE 4
MANOVA RESULTS OF LANGUAGE AND DISCOURSE VARIABLES OF EDUCATION (HUMANITIES & MEDICINE)

| Feature (language variable) | Examples | F | $\eta^2$ |
|---|---|---|---|
| Type-Token Ratio (TTR) | | 1129.50** | 0.204 |
| Arousal words | feel, excite | 899.33** | 0.169 |
| Number of words | | 428.07** | 0.088 |
| First person singular | I, me | 352.18** | 0.074 |
| Positive sentiment | love, like, excite | 302.02** | 0.064 |
| Number of words/sentence | | 257.66** | 0.055 |
| Pedagogical | lecturer, grade, video | 238.06** | 0.051 |
| Question bi-gram | Can someone | 236.39** | 0.051 |
| Third person pronoun | he, she, them | 212.18** | 0.046 |
| Number of sentences | | 171.05** | 0.037 |
| Negations | not, couldn't, do not | 160.91** | 0.035 |
| Neutral sentiments | | 150.42** | 0.033 |
| Gratitude, politeness | appreciate, please, sorry | 126.93** | 0.028 |
| Question mark | ? | 86.39** | 0.019 |
| Demonstrative determiners | this, these | 61.31** | 0.014 |
| Positive emotion | pleasure, enjoyment | 42.98** | 0.010 |
| Negative sentiments | unable, awful, stress | 33.79** | 0.008 |
| Negative emotion | anger, fear, disgust | 30.09** | 0.007 |
| Problem solving | issue, question, doubt | 27.21** | 0.006 |
| Second person pronouns | you, your | 22.04** | 0.005 |
| Future words | will, might, would | 16.29** | 0.004 |

**$p<0.001$; Note: significant language variables are similar in Humanities and Medicine domains. However, F and $\eta^2$ are slightly different. Language variables are ranked based on their effect size.

According to our first research question on "*what language and discourse features demonstrate a significant difference between confused and non-confused posts?*", Table 4 provides indications on which linguistic features are significantly different between confused and non-confused categories within all three domains. Additionally, Table 5 reports remaining significant language variables within the Humanities and Medicine domain. A comprehensive interpretation of predictive language and discourse variables are included in Section 5.A.

TABLE 5
MANOVA RESULTS OF LANGUAGE AND DISCOURSE VARIABLES OF HUMANITIES (& MEDICINE)

| Feature (language variable) | Examples | F | $\eta^2$ |
|---|---|---|---|
| Number of pronouns | | 90.96** | 0.028 |
| Question stem | who, how, what | 61.03** | 0.019 |
| Confusion expressions | don't understand | 49.22** | 0.015 |
| Incomplete expressions | incomplete, missing | 37.93** | 0.012 |
| Opinion | I think, probably | 17.84** | 0.006 |

**$p<0.001$; Note: significant language variables are similar in Medicine domain. However, F and $\eta^2$ are slightly different. Language variables are ranked based on their effect size.

## B. Confusion classification

Building on the findings from the statistical analysis, we construct a set of binary classifiers for confusion detection, addressing RQ2 '*Can we build machine learning models for confusion classification solely based on linguistic features?*'. We utilise predictive 175, 153, and 118 features from the Education, Humanities, and Medicine domains respectively to build classifiers.

Prior to building models, the datasets that have imbalanced class distributions (e.g. Table 2) are adjusted using the popular oversampling technique namely Synthetic Minority Oversampling Technique (SMOTE) [36]. In other words, synthetic instances were created for the classes that have less number of instances to balance the ratio (approximately 50% - 50%) between the two classes. To avoid overoptimistic performance of the classifiers, they were trained using the synthetic instances and tested with the realistic imbalanced dataset.

We implement the following classifiers using Weka[5]: Naïve Bayes, Support Vector Machine (SVM), Sequential Minimal Optimisation (SMO), Random Forest, Ada Boost M1, Simple Logistic Regression, and Logistic Regression. To choose the best classifier, we measure the classifier performance using 10-Fold Cross-Validation. Random Forest is selected as the best performing model in all three domains (Note: our confusion classifiers will be shared publicly upon acceptance of the manuscript for publication).

We present the performance of our Random Forest models with neutral posts (model 2) and without neutral posts (model 1) and three baselines in Fig. 1 (see data table in the Appendix - Table A). When calculating the F1 difference (7th column in Table A), we utilise either model 1 or 2 based on the corresponding datasets of baselines. If the underlying dataset is not clear in the publication, we compared with our default model (i.e. model 2) (Note: we acknowledge the reuse of comparison data from Zeng *et al.* [23]).

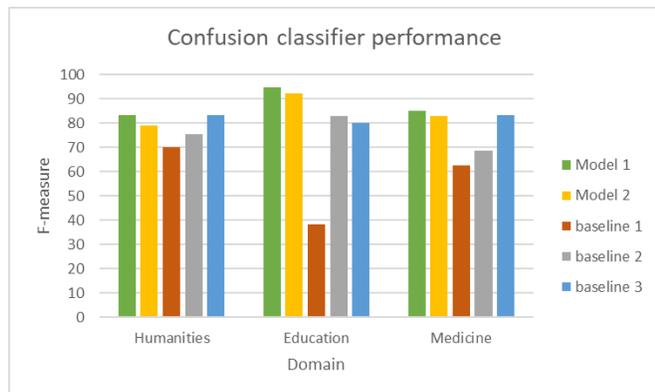

Fig. 1. Confusion classifier performance (F-measure) using 10-fold cross-validation; Model 1 (linguistic) – without neutral posts, Model 2 (linguistic) – with neutral posts, baseline1 (unigram + community) [8], baseline 2 (unigram) [25], baseline 3 (unigram + community + content) [23]

With over 83.1% (model 1) and 79% (model 2) of F-measure in all three domains, it is evident that the *linguistic-only* features are highly effective in confusion classification, addressing our second research question. Our model outperforms the baselines that utilised an identical dataset (i.e. Stanford MOOC posts [26]) by a substantial margin (see 'F1 difference' in Table A).

Further comparisons such as replicability, practicality are restricted since baselines do not report the measures like the total amount of features used or classification speed. In terms of the sample, baseline 2 considered neutral posts as confused while baseline 3 omitted neutral posts from the analysis.

Additionally, our model demonstrates promising results compared to the linguistic- or question-only models built using a different dataset by Yang *et al.* [14]. They obtained an accuracy of 64% for linguistic and 68% for question model in Algebra course and 59% for linguistic and 65% for question model in Micro Economics.

---

[5] https://www.cs.waikato.ac.nz/ml/weka/

### C. Cross-domain validation

In order to address our third research question on '*Can we successfully validate the linguistic-based machine learning models for confusion classification on unseen domains?*', we trained our classifiers in one domain and tested in other domains (Table B & Fig. 2). In this study, we situate our work with previous baseline 3 [23] as other baselines have not reported cross-domain validation.

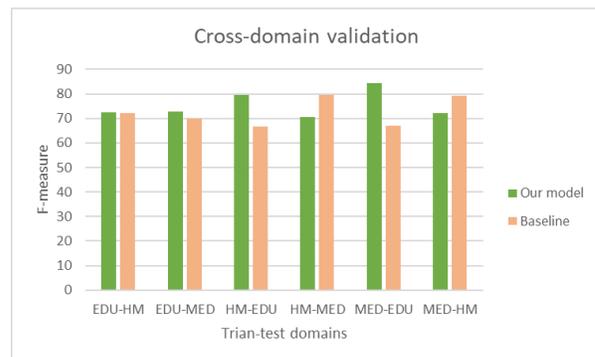

Fig. 2. Cross-domain validation (F-measure); Our model - linguistic, baseline [23] - unigram + content + community

According to the Fig. 2, we obtained 'good' performance with over 70.7 of F-measure for all domain pairs (i.e. training on one domain and testing on another domain), addressing RQ3. Thus, we confirm that *linguistic-only* model is effective in confusion classification independent from the domain and community-related features. Our classifier outperforms the baseline in all instances except Humanities-Medicine pair (Fig. 2).

## V. DISCUSSION

Firstly, we provide an interpretation of our findings. Then, we list our contributions followed by limitations. Finally, we summarise the implications for research and practice.

### A. Interpretation of results

This section summarises the highly predictive language features for confusion identification, situating with the implications from related works. We organise the discussion around 1) features extracted from NLP tools, 2) direct confusion, 3) indirect confusion expressions, 4) cross-domain incompatible features, 4) classifier performance, and 5) cross-domain validation, enabling the alignment with our methodology. The first four subsubsections lead to the interpretation of predictive language and discourse features whereas the last two subsections lead to the interpretation of classifier performance.

### 1) Features extracted from NLP tools

Table 4 demonstrates *Type-Token Ratio (TTR)* [37] as the highest predictive feature for identifying confusion in all three domains. TTR, the ratio between unique words (types) and the total number of words (tokens) in a sentence is used to measure lexical diversity. When TTR equals 1, each word occurs only once in the sentence. The mean difference of TTR between two categories (within the Education domain, confused class: M = 0.85, SD = 0.10, n = 3153; non-confused class: M = 0.73, SD = 0.08, n = 6690) implies confused learners are more likely write lexically diverse posts, i.e. higher TTR values than non-confused learners. This observed phenomenon will be further explained using the 'post length' below since McCarthy and Jarvis [38] suggests *"when the number of word tokens increases, there is a lower likelihood of those words being unique"*.

Descriptive features like *number of words, number of words per sentence,* and *number of sentences* are highly predictive for confusion classification in all three domains. This outcome is consistent with previous findings by Zeng *et al.* [23] which suggests that *post length* is highly predictive for confusion classification. However, our results are contradictory to the finding above of *lexical diversity*. The mean difference of post length (a.k.a. number of words) between confused (M = 51.46, SD = 47.65, n = 6690) and non-confused (M = 106.09, SD = 63.73, n = 3153) category suggests that non-confused learners tend to write longer posts. This outcome can be used to explain the lower lexical diversity (TTR) within the non-confused category [38].

Our results demonstrate 'arousal words' (e.g. *feel, excite*) and 'positive sentiments' (e.g. *love, enjoy*) which indicate excitement as highly predictive features for identifying confusion in all three domains. Based on the mean differences between categories, learners who use positive sentiments or arousal words are less likely demonstrate confusion state. SÉANCE utilised EmoLex lexicons list [39], VADER (Valence Aware Dictionary and sEntiment Reasoner) [40] as a thesaurus to calculate scores for sentiment and emotion metrics. Our findings are consistent with previous studies, which suggest learners' whose discussion forums include positive sentiments are positively correlated with MOOC success, motivation, and performance [41-43].

*"I really enjoyed reading first ten pages specially about surface of triangle [...]"*
[**positive & non-confused**]

However, the use of positive sentiment for confusion classification is occasionally challenging as unigram features like *good, like, great* are common in non-confused posts while bigram features like *(would) be good, would like* that convey 'politeness' or 'gratitude' (Table 4) is often correspond to asking help by confused learners. Therefore, features related to politeness or gratitude can be presented as related to 'indirect confusion' (refer to Section 5.A.3).

*"It would be good to have a response from the course provider if it is really a wrong spelling [...]"*
[**politeness & confused**]

Further, our results confirm that learners who tend to use positive emotions in their language such as *pleasure, enjoyment,* and *enthusiasm* are likely not confused.

Conversely, posts that are annotated as confused often include negative sentiments (e.g. *unable, disappoint*), negative emotions (e.g. *anger, fear, disgust, anxiety, sad*), and negations (e.g. *do not, not*) than others. For instance, the mean difference of negations between confused (M = 0.02, SD = 0.002, n = 6690) and non-confused (M = 0.01, SD = 0.01, n = 3153) category within the Education domain demonstrate significant difference with larger effect size (F = 160.91, $p<0.001$, $\eta^2 = 0.035$). Although, negative sentiments and emotions demonstrate significant difference between confused and non-confused categories, their effect was not large (see Table 4).

*"I was unable to install Deducer to Mac?? Multiple attempts [....]. Any alternatives/ suggestions?"*
[**negative sentiments & confused**]

*"Isn't there anyway they can verify [.....] but this online class had loading error? ☹ This is so sad it affects that grade so bad! [...]"*
[**negative emotion & confused**]

Intuitively, posts that use 'neutral' sentiments might indicate a non-confused state as no emotions are implied. However, our findings drawn from the statistical analysis are not consistent with this intuition. The results demonstrate that the posts with neutral sentiments highly likely belong to a confused category in all three domains. For example, the mean difference of 'neutral sentiment' between confused (M = 0.85, SD = 0.16, n = 6690) and non-confused category (M = 0.80, SD = 0.78, n = 3153) within the Education domain is statistically significant (Table 4). This is an important implication drawn from our study. Learners might be confused even if their language does not express negativity.

Thus, although the use of sentiments, emotions, and negations are effective for confusion identification, the future models should not solely depend on these features as our study shows that confusion can also be expressed neutrally. Further, according to our findings, positive sentiments cannot be solely used to distinguish non-confused learners since positive bigrams (e.g. *would* like), politeness- or gratitude-related features imply confusion state. Moreover, the use of emotional words are not very frequent in our corpus and therefore, they may not produce strong predictors for confusion classification.

*2) Direct confusion expressions*

'Question', as expressed directly, is a strong variable to distinguish confused and non-confused posts. Among them, 'question bigrams' such as *can someone, what if,* and *any suggestion* show higher statistical significance (F = 239.39, $p < 0.001$, $\eta^2 = 0.051$) than 'question mark' (F = 86.39, $p < 0.001$, $\eta^2 = 0.019$) in all three domains. These findings are consistent with earlier works [8, 14, 23]. A correlational study by Agrawal *et al.* [8] found that confusion is positively correlated with the question variable. Through a feature ablation study, Zeng *et al.* [23] demonstrated that removing

'question mark' as a feature in their model decreases performance (F-measure) by 11.14, 2.12, and 15.61 in the Humanities, Education, and Medicine domain respectively. Yang et al. [14] demonstrated that 'question-related' features are the second best individual feature set after 'bag of words' features for confusion classification.

> *"This is the first module that made me <u>confused</u>. We need to [...], simulation software works<u>?</u>"*
> [**question mark & confusion expression**]

Features related to 'problem-solving' (e.g. *error, issue, solution, quiz*) are significant in confusion identification in all three domains (Table 4). This suggests confused learners might not explicitly express confusion, but their language is influenced by their problem-solving intentions.

> *"<u>Is there</u> any way to lodge a <u>complaint</u> about the <u>quiz</u> this <u>week</u>? There were <u>no</u> <u>instructions</u> and <u>I</u> got a few <u>answers</u> <u>wrong</u> because <u>I</u> thought only one <u>choice</u> was allowed [..]"*
> [**problem-solving & confusion**]

*3) Indirect confusion expression*

We observed interesting patterns with the use of 'pronouns'. Our results confirmed the use of third-person pronouns (e.g. *he, they*) and demonstrative determiners (e.g. *these, those*) are significantly higher (in Education & with the use of third-person pronouns, Confused – M = 0.01, SD = 0.02, n = 6690; non-confused – M = 0.03, SD = 0.03, n = 3153) among non-confused learners in all three domains (Table 4). This suggests that the learners who do not exhibit confused state often refer to the third party in their discussions instead of relating to themselves (e.g. *I, my*).

Second-person pronouns, predominantly the use of the term 'you', were significantly higher among non-confused posts in all three domains (in Medicine, confused – M = 0.01, SD = 0.02, n = 8324; non-confused – M = 0.02, SD = 0.02, n = 1563). According to empirical evidence [7], confused learners tend to initiate threads and therefore, the use of the term 'you', which implies a conversational dialogue is uncommon in a new thread. D'Mello & Graesser [7] also found that lower use of second person pronouns by the student coupled with increased use of 'future tense words' by the tutor is predictive for confusion classification in ITSs.

Our findings on 'future words' (e.g. *will, might*) which demonstrate that future words are statistically significant (F = 16.29, *p < 0.001*, $\eta^2$ = 0.004) in confusion identification are consistent with D'Mello & Graesser [7].

Conversely, first-person singular pronouns (e.g. *I, my*), as one of the most predictive features (Table 4), demonstrate significantly higher use in confused posts, indicating that they often focus on self-reflection. Previous work by Wen et al. [41] suggests *"first-person pronouns indicate that the user can relate the discussion to self effectively"*. According to D'Mello & Graesser [7], first-person singular pronouns also linked with negative emotions. This behaviour within our corpus is therefore considered as highly predictive and desired, as it provides valuables insights for identifying confusion.

Moreover, 'number of pronouns' in a post also provides a predictive indicator of confusion. According to Table 5, the mean difference of number of pronouns is statistically significant between confused and non-confused classes within the Humanities and Medicine domains. Although not included in Table 4, our analysis also demonstrated that use of number of pronouns is higher among confused learners within the Education domain, however, the effect is small (F = 10.96, *p < 0.05*, $\eta^2$ = 0.002).

> *"<u>I'm</u> doing <u>my</u> best to understand medical statistics but <u>I</u> have been getting very low scores on <u>my</u> latest homework assignments [...]"*
> [**first-person pronouns & indirect confusion**]

Interestingly, 'pedagogical' features, i.e. lexicons related to introduce teaching and learning strategies such as *assessment, quiz, video,* and *lecture* are statistically significant in identifying confusion in all three domains. This suggests that learners' confusion often associated with pedagogical strategies [44].

> "I went through and completed all of my <u>tasks</u> and watched all of the <u>videos</u> for entry 1 yet it shows that I made <u>no</u> <u>progress</u> on the <u>course</u> <u>progress page</u>"
> [**pedagogical features & indirect confusion**].

Finally, as discussed in Section 5.A.1, features related to 'gratitude/politeness' is also considered as associated with indirect confusion.

*4) Cross-domain incompatible features*

We obtained inconsistent results for features such as question stem, direct confusion expressions, incompleteness-related expressions, and opinion words across domains. These features are primarily predictive in the Humanities and Medicine domains but not in the Education domain.

Although question mark and question bi-gram demonstrate significant results across domains, 'question stem' (e.g. *what, why*) does not demonstrate significant results within the Education domain (F = 0.33, *p > 0.05*, $\eta^2$ = 0) but Humanities and Medicine domains (Table 5). This suggests that confused learners might not explicitly start a post with a question stem, suggesting the importance of incorporating other question-related indicators like question bi-grams, question mark upon the development of generalised confusion classfiers.

Confusion features that express directly (e.g. *confuse, frustrate*) or use expressions related to incompleteness, i.e. uncertainty, doubt and vagueness (e.g. *missing, nothing*) [9] were significantly different between groups in all domains excepts Education. For instance, Medicine domain demonstrates statistically significant mean difference of 'confusion expressions' between confused (M = 0.01, SD = 0.03, n = 8324) and non-confused (M = 0.003, SD =0.008, n = 1563) posts. Conversely, the mean difference is not significant in the Education domain (F = 1.08, $p > 0.05$, $\eta^2 = 0$). However, the effect of confusion or incomplete expressions were also small in the Humanities and Medicine domains (Table 5).

> *"Hi all, I have a doubt. Please, could any of you explain me this point that I don't understand [...]"*
> [**Confusion expression**]

> *"I really thought I understood what is going on but I am still puzzled [...] I don't understand what I am missing."*
> [**Incomplete expression**]

Even though the effect is small, 'opinion' (or point-of-view) features (e.g. *I think, I believe, personally, probably*) demonstrate statistically significant results in the Humanities (F = 17.84, $p < 0.001$, $\eta^2 = 0.006$) and Medicine (F = 4.76, $p < 0.05$, $\eta^2 = 0.003$) domains but not in the Education domain. Non-confused learners tend to use opinion phrases significantly higher than confused learners within the Humanities and Medicine domains. In a related study, Elouazizi [45] demonstrated cognitively disengaged learners will have a low frequency of 'opinion' language while the cognitively engaged learners will demonstrate the high use of opinion components in their writing.

> *"I'm a fellow student but from what I've read from the emails introducing the course, I believe I can answer your question [...]"*
> [**opinion & non-confused**]

Thus, due to the demonstrated incompatibility, through analysis, we suggest that question stem, confusion expressions, incompleteness-related expressions, and opinion phrases are not suitable for building cross-domain compatible confusion classifiers as they are domain dependent.

*5) Confusion classification*

In comparison to previous best performing models by Zeng *et al.* [23], our best models achieved 83.1, 94.5, and 85.1 of F-measure for the Humanities, Education, and Medicine respectively. This level of classifier performance is considered as good and satisfactory for text classification [46].

However, with the achieved performance, we ask ourselves *how good is good enough?*. For the task of confusion classification, we prefer higher recall over precision. This means classifier should reduce its false negatives (i.e. confused posts being classified as non-confused), enabling the instructors to identify confused posts early prior to the transition of learners' confusion state from frustration, and eventually, to boredom. Conversely, having more false positives (i.e. non-confused posts being classified as confused) which leads to low precision is relatively acceptable since our goal is to minimise the chance of bewilderment. Accordingly, we achieved a high level of recall, i.e. 82.8, 94.8, and 85.3 for the Humanities, Education, and Medicine respectively (Fig. 3).

Thus, with respect to our second research question, building confusion classifiers solely based on linguistic features with high F-measure, recall, and accuracy (Table A) is supported by our study within the Humanities, Education, and Medicine domains. Overall, our classifier performed well within the Education domain with an F-measure of 94.5, and the F-measure difference was greater than 9.24 for any previous works built using an identical dataset (Table A).

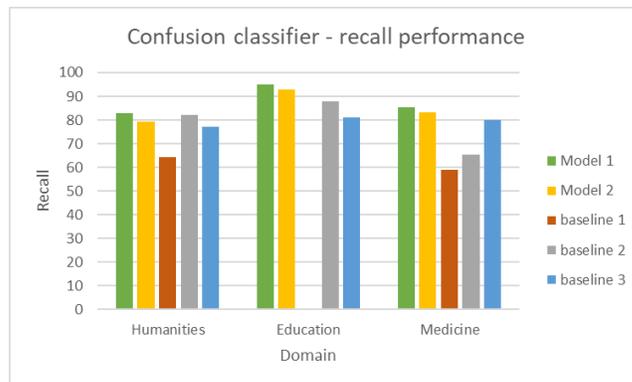

Fig. 3. Confusion classifier performance (Recall) using 10-fold cross-validation; Model 1 (linguistic) – without neutral posts, Model 2 (linguistic) – with neutral posts, baseline1 (unigram + community) [8], baseline 2 (unigram) [25], baseline 3 (unigram + community + content) [23]

*6) Cross-domain validation*

We achieved a satisfactory performance of over 70.7 F-measure [46] for all domain pairs in our cross-domain validation study. Yet, training on Education domain and testing on other domains or vice versa outperformed Zeng *et al.* [23], suggesting that the classifier associated with the Education data is the best performing model for unseen data. The previous best model for cross-domain confusion classification by Zeng *et al.* [23] performed better than us only in the Medicine-Humanities pair (Fig. 2).

According to Table 4 & 5, Medicine and Humanities domain pair has consistent predictive variables whereas Education domain does not demonstrate significant results for the variables such as question stem, confusion and incompletion-related expressions, and opinion phrases (see more information in 'cross-domain incompatible features' under Section 5.A.4) which might explain the observed phenomenon.

*B. Contributions*

We make the following contributions:
- Language analysis of approximately 30,000 discussion

posts to propose a novel set of predictive linguistic features for confusion classification. The 'text-only' aspect of our work provides easy replication. Also, as our approach is independent of community-related features, the 'early' identification of confusion is feasible without the influence from the community (i.e. view, votes), enabling the usability within a personalised context.
- Our study has drawn interesting implications for linguistic feature selection for confusion classification (see Section 5.D).
- We design and evaluate confusion classifiers that effectively classify confused posts with a good F-measure over 79% and 70.7% within three domains (Education, Medicine, and Humanities) and across domains respectively, indicating that these domain independent linguistic features have potentials to be adopted in any online learning platform. Our classifiers outperform baselines in every instance except cross-domain validation of Humanities-Medicine pair.
- Due to the relatively low amount of features used for training (approximately 150) and high speed (19s on average) using a laptop with the specifications as Intel Core i7-8850H 2.6 GHz CPU, 16GB RAM, our classifiers are lightweight and efficient to be implemented in the real-time environment.

*C. Limitations*

Due to the large sample sizes in each course (n ~ 10000), the effect ($\eta^2$) of language variables were too small (Table 4 & 5). This behaviour was expected. Therefore, these variables do not make any sense individually due to low effect, but using a combined set of features is desirable for the classification task.

We categorised neutral posts into the confused class to reduce false negatives. However, there is a practical problem if all neutral posts being classified as confused, providing challenges for instructors to identify which posts need urgent interventions. Therefore, our model can be further improved as a multi-class classification problem (confused, neutral, and non-confused), enabling instructors to allocate priority to the confused class.

Our current model introduces some bias towards lengthy posts as they include more features than short posts. However, our motivation is to correctly classify every posts, decreasing false negatives. To achieve this, we have to consider every single posts irrespective of their size.

*D. Implications for research/practice*

Our study urged the importance of linguistic clues for automated confusion classification. As a guide for research/practice, we summarise the indicators of confusion classification in Table 6. The second column in Table 6 lists the variables that are already contributed from the related literature for confusion identification [7, 8, 14, 23, 41-43] or the variables that directly expresses confusion (e.g. negations, confusion expressions). The third column of the table lists the novel variables contributed from this study. We organise these variables based on the majority class (i.e. either confused or non-confused).

TABLE 6
SUMMARY OF FINDINGS FOR RESEARCH/PRACTICE

|  | Direct/Already known | Indirect/New |
|---|---|---|
| Confused | negative sentiment/emotions<br>negations<br>question mark<br>question stem*<br>first person pronouns<br>confusion expression*<br>incomplete expression* | lexical diversity (TTR)<br>politeness/gratitude<br>neutral sentiments<br>problem-solving<br>number of pronouns<br>pedagogical<br>question bigram |
| Non-confused | number of words<br>number of sentences<br>number of words/sentence<br>positive sentiment/emotions<br>second person pronouns<br>future words | arousal words<br>third person pronouns<br>demonstrative determiners<br>opinion* |

*cross-domain incompatible variables

## VI. CONCLUSION AND FUTURE WORKS

Learners in massive classrooms are quite isolated and need more support. Therefore, early detection of confusion has potential benefits including rectifying time-critical misconceptions and prevention of transitioning the learners' emotional state from confusion to frustration/boredom that leads to course attrition. Due to the dependability only on the language expressed by the individual learner (i.e. text-only aspect) and validity across domains, the model proposed in this study has potential to adopt to other platforms such as online communities (e.g.Stackoverflow), ITSs, LMS, and within course evaluation studies etc.

In future works, we not only apply state-based data but intend to incorporate trait-based data (e.g. personality, confidence) and demographics (e.g. age, gender) of individuals to provide personalised interventions such as supplementary resources [8], post responses, further explanations for confused learners, preventing them to enter the maximum level of adverse confusion.

ACKNOWLEDGEMENT

Authors would like to acknowledge the authors of Stanford MOOC posts dataset.